# Impact of Non-universal $Z'$ in the Lepton Flavour Violating $B(B_s) \to K^*(\varphi) l_1^- l_2^+$ Decays

**S. Biswas[1], S. Mahata[2], A. Biswas[3] and S. Sahoo[4]**

[1,2,4]Department of Physics, National Institute of Technology Durgapur
Durgapur-713209, West Bengal, India

[3]Department of Physics, Banwarilal Bholatia College,
Asansol – 713303, West Bengal, India

[1]E-mail: getswagata92@gmail.com, [4]E-mail: sukadevsahoo@yahoo.com

**Abstract**

In recent years, lepton flavour violating (LFV) decays are one of the most trending topics to probe new physics (NP). The latest results of LHCb have motivated us to study the LFV decays through $b \to s$ transition. The branching ratios, forward backward asymmetries and longitudinal polarization fractions of $B(B_s) \to K^*(\varphi) l_1 l_2$ decays are studied in non-universal $Z'$ model. Here, we have structured the four-fold angular distribution of the decays in terms of transversity amplitudes. The variation of the observables in whole kinematic region shows the sensitivity of NP. The observables estimated in this work are very intriguing and might lead a new track towards NP in near future.

## I.    Introduction

Inspired by the various LHCb results of lepton flavour violation on $b \to s$ transition, we study the lepton flavour violating $B(B_s) \to K^*(\varphi) l_1 l_2$ decays in terms of transversity amplitudes in non-universal $Z'$ model. Recently, some B meson decay measurements disagree with the prediction of the standard model (SM) known as B flavour anomalies. Some of them are: (i) neutral-current anomalies involved by $b \to s l^+ l^-$ transition, (ii) charged-current anomalies involved by $b \to c l^- \bar{\nu}_l$ transition, (iii) lepton flavour violation (LFV). The LFV decays are highly suppressed in the SM as the predicted values in the SM lie far below recent experimental results. For example, in ref. [1], branching ratios for $B_d \to \tau^\pm \mu^\mp$ and $B_s \to \tau^\pm \mu^\mp$ decays are calculated in the SM of order $10^{-54}$ whereas Babar and LHCb has constrained the values at the order of $10^{-5}$ with 90% and 95% C. L in refs. [2, 3]. The SM theory conserves the generation lepton number of electroweak interactions but the lepton flavour violation has been traced by the observations of neutrino oscillation. The mismatch between weak and mass eigenstates of neutrinos causes the neutrino oscillation as well as the mixing between various generation of leptons via charged current interactions of W boson [4]. It is well-known that the principle for the flavour changing neutral current (FCNC) transitions of lepton sector in LFV decays could be similar to that for the quark sector. The FCNC transitions in the lepton sector which are due to the mixing in the charged current (CC) interactions with the left-handed W boson and light neutrinos are very small because they are suppressed by powers of $m_\nu^2/m_W^2$. There are many other observables to examine lepton flavour universality (LFU) with FCNC. After the discovery of anomaly of the observable $P_5'$, a discrepancy in the measurement of $R_{K^{(*)}} = \frac{Br(B \to K^{(*)} \mu^+ \mu^-)}{Br(B \to K^{(*)} e^+ e^-)}$ was found. It is recently announced that $(R_K)_{new} = 0.846^{+0.060+0.016}_{-0.054-0.014}, 1 \leq q^2 \leq 6 \text{ GeV}^2$ [5] (where $q^2$ is the momentum transfer term) which is lesser than the SM value $(R_K)_{SM} = 1.00 \pm 0.01$ [6] by $2.5\sigma$ deviation. The value of $R_{K^*}$ is observed at LHCb [7]. Belle experiment has also constrained $R_K$ and $R_{K^*}$ values which are nearer to the SM [8] prediction but with greater uncertainties. The lepton flavour universality violation test is performed for the charged-current transitions in 2012 for the first time with the observable $R_D$ and $R_{D^*}$. The colliders of the B factories: Belle [9-11], BaBar [12] and LHCb [13] have set their values of $R_{D^*}$. Recently, Belle has announced the values as [14]: $R_D = 0.307 \pm 0.37 \pm 0.016$ and $R_{D^*} = 0.283 \pm 0.018 \pm 0.14$. All these values are more than the SM values given in ref. [15] and ref. [16] by $2.3\sigma$ and $3.4\sigma$ deviations respectively.

Although the SM has provided large number of predictions and most of them are experimentally confirmed, it provides many unanswered questions in terms of the above spoken anomalies. So it is clear that the SM is not complete and we need the physics beyond the SM. Various theoretical models are proposed to explain the experimental tensions of LFV processes within current sensitivity. The LFV decays, such as $\tau \to 3\mu, \mu \to 3e, l \to l'M$ (where $l, l'$ are leptons of different flavours and $M$ reprsents meson) and radiative decays $\mu \to e\gamma$ etc, all are analysed in different NP models though they are not evidently seen experimentally but their experimental upper limits exist. The LFV decays are successfully illustrated with the effect of FCNC mediated $Z$ boson [4, 17], non-universal $Z'$ boson [18], leptoquarks [19] and other NP models and also in model independent way [20].

In this work, we study the differential branching fractions, forward backward asymmetries and longitudinal polarization fractions of LFV decays $B \to K^* l_1 l_2$ and $B_s \to \varphi l_1 l_2$ induced by the quark level transition b$\to s l_1 l_2$ in $Z'$ model where $l_1$ and $l_2$ are charged leptons of different flavours. We constrain the NP couplings using several experimental upper limits.

Here, we consider non-universal $Z'$ model [21-25]. In this model, the NP contributes at tree level via $Z'$-mediated flavour changing $b \to s$ decays where $Z'$ associates to the quark sector $\bar{s}b$ as well as to the leptonic sector $l_1^- l_2^+$. Conventionally in different grand unified theories (GUTs) the $Z'$ mass is considered differently as it is not unveiled till now. So the $Z'$ mass is restricted by various experiments and detectors constraining its upper and lower limit. The model-dependent lower bound is set at 500 GeV [26-28]. The mass of $Z'$ boson is predicted by Sahoo et al. in the range 1352-1665 GeV from $B_q - \bar{B}_q$ mixing [29] and the LHC also constrained the $Z'$ mass, the $Z - Z'$ mixing angle and the coupling of extra $U(1)$ gauge group [30-32]. Lower limit of $Z'$ mass is set as 2.42 TeV [30] and 4.1 TeV [33] for sequential standard model (SSM) and for $E_6$ motivated $Z'_\chi$ by the ATLAS collaboration respectively. The CMS collaboration has also set the lower bound of $M_{Z'}$ as 4.5 TeV for sequential standard model (SSM) and 3.9 TeV for superstring-inspired model [31]. Bandopadhyay et al. have constrained as $M_{Z'} > 4.4$ TeV using the recent Drell-Yan data of the LHC [32]. Recently [34], the mass difference of $B_s$ meson is studied in extended standard model and the upper limit of $Z'$ mass is constrained as 9 TeV. The ATLAS collaboration has set the lower limit of $Z'$ mass at 4.5 TeV and 5.1 TeV for $Z'_\psi$ and $Z'_{SSM}$ at 95% C. L. [35]. In this paper, we have taken the $Z'$ mass in TeV range.

The paper is arranged as follows: The effective Hamiltonian for LFV decays is discussed in Sec. II. The kinematics of the decays, detail of transversity amplitudes and the description of the observables are illustrated in Sec. III. The NP is introduced in the next section, i.e. the contribution of non-universal $Z'$ model is incorporated in Sec. IV. We have presented the numerical analysis in Sec. V. And finally we have concluded the findings in Sec. VI.

## II. Effective Hamiltonian of $b \to s l_1^- l_2^+$

In this section, we structure the effective Hamiltonian for the lepton flavour violating $b \to s l_1^- l_2^+$ transition. The leptons $l_1^-$ and $l_2^+$ are considered of the same flavour $l$ in the SM but in BSM physics the NP particle $Z'$ will couple differently with leptons of different families. The structured Hamiltonian is represented as follows [36-38]

$$\mathcal{H}^{eff} = -\frac{G_F \alpha}{2\sqrt{2}\pi} V_{tb} V_{ts}^* \sum_{r=9,10} C_r^{NP} O_r + h.c., \tag{1}$$

where $G_F$ represents the Fermi coupling constant, $\alpha$ represents the electromagnetic coupling constant. The primed parts are the Wilson Coefficients containing NP contributions. The CKM matrix elements $V_{tb} V_{ts}^*$ are introduced in the Hamiltonian due to the virtual effects induced by $t\bar{t}$ loops. It is to be noted that the LFV decays always occur at tree level in this $Z'$ model; therefore the NP should contribute in the fashion where $t\bar{t}$ loops get cancelled. Moreover, there is an electromagnetic operator $O_7$ in the SM for $b \to sll$ transition. Non-universal $Z'$ model is basically sensitive to the semileptonic current operators $O_9$ and $O_{10}$ involving NP contributions in $C_9^{NP}$ and $C_{10}^{NP}$ [38-40]. Here,

$$\begin{aligned} O_9 &= [\bar{s}\gamma_\mu(1-\gamma_5)b][\bar{l}_1\gamma^\mu l_2] \\ O_{10} &= [\bar{s}\gamma_\mu(1-\gamma_5)b][\bar{l}_1\gamma^\mu \gamma_5 l_2], \end{aligned} \tag{2}$$

## III. The $B \to K^* l_1 l_2$ and $B_s \to \varphi l_1 l_2$ decays

The $B \to K^* l_1 l_2$ and $B_s \to \varphi l_1 l_2$ decays proceed via $B \to K^*(\to K^- \pi^+) l_1 l_2$ and $B_s \to \varphi(\to K\bar{K}) l_1 l_2$. We have adopted the kinematics from ref. [41]. Our angular conventions are taken as [37], i) $\theta_l$, the angle made by $l_1$ lepton with z axis in the dilepton rest frame, ii) $\theta_K$, the angle made by $K^-$ with decay axis- z, iii) $\phi$, the angle between the planes spanned by $K\pi$ ($K\bar{K}$) and $l_1^- l_2^+$. The leptonic and hadronic four-vectors can be defined as, $q^\mu = (q_0, 0, 0, q_z)$ and $k^\mu = (k_0, 0, 0, -q_z)$, where

$$q_0 = \frac{q^2 + m_B^2 - m_V^2}{2m_B}, \quad k_0 = \frac{m_B^2 + m_V^2 - q^2}{2m_B}, \quad q_z = \frac{\sqrt{\lambda(\sqrt{q^2}, m_B, m_V)}}{2m_B}.$$

Here $V$ represents the vector mesons $K^*(\to K^-\pi^+)$ and $\varphi(\to K\bar{K})$ meson. The leptonic four-vectors can be read as,

$$p_1^\mu = (E_\alpha, |p_l|\sin\theta_l\cos\phi, -|p_l|\sin\theta_l\sin\phi, |p_l|\cos\theta_l),$$
$$p_2^\mu = (E_\beta, -|p_l|\sin\theta_l\cos\phi, |p_l|\sin\theta_l\sin\phi, -|p_l|\cos\theta_l),$$

where

$$E_1 = \frac{q^2 + m_1^2 - m_2^2}{2\sqrt{q^2}}, \quad E_2 = \frac{q^2 - m_1^2 + m_2^2}{2\sqrt{q^2}}, \quad |p_l| = \frac{\sqrt{\lambda(q^2, m_1^2, m_2^2)}}{2m_B}.$$

In the similar manner, we can write in the $K^*(\varphi)$ rest frame

$$p_K^\mu = (E_K, -|p_K|\sin\theta_K, 0, -|p_K|\cos\theta_K),$$
$$p_{\pi(K)}^\mu = (E_{\pi(K)}, +|p_K|\sin\theta_K, 0, +|p_K|\cos\theta_K),$$

where $E_K$, $E_{\pi(K)}$ and $|p_K|$ are defined by the similar type expressions.

The full angular distribution of the above decay can be read as,

$$\frac{d^4\mathcal{B}}{dq^2 d\cos\theta_l d\cos\theta_K d\phi} = \frac{9}{32\pi} I(q^2, \theta_l, \theta_K, \phi), \tag{3}$$

with
$I(q^2, \theta_l, \theta_K, \phi)$
$$= I_1^s(q^2)\sin^2\theta_K + I_1^c(q^2)\cos^2\theta_K + [I_2^s(q^2)\sin^2\theta_K + I_2^c(q^2)\cos^2\theta_K]\cos2\theta_l$$
$$+ I_3(q^2)\sin^2\theta_K\sin^2\theta_l\cos2\phi + I_4(q^2)\sin2\theta_K\sin2\theta_l\cos\phi$$
$$+ I_5(q^2)\sin2\theta_K\sin\theta_l\cos\phi + [I_6^s(q^2)\sin^2\theta_K + I_6^c(q^2)\cos^2\theta_K]\cos\theta_l$$
$$+ I_7(q^2)\sin2\theta_K\sin\theta_l\sin\phi + I_8(q^2)\sin2\theta_K\sin2\theta_l\sin\phi$$
$$+ I_9(q^2)\sin^2\theta_K\sin^2\theta_l\sin2\phi. \tag{4}$$

Proceeding with these considerations, we have found the differential decay rate of the above decay as,

$$\frac{d\mathcal{B}}{dq^2} = \frac{1}{4}[3I_1^c(q^2) + 6I_1^s(q^2) - I_2^c(q^2) - 2I_2^s(q^2)]. \tag{5}$$

Another interesting tool for exploring new physics in these lepton flavour violating decays is the forward-backward asymmetry which is given below,

$$A_{FB}(q^2) = \frac{\int_0^1 \frac{d^2\mathcal{B}}{dq^2 d\cos\theta_l} d\cos\theta_l - \int_{-1}^0 \frac{d^2\mathcal{B}}{dq^2 d\cos\theta_l} d\cos\theta_l}{\int_0^1 \frac{d^2\mathcal{B}}{dq^2 d\cos\theta_l} d\cos\theta_l + \int_{-1}^0 \frac{d^2\mathcal{B}}{dq^2 d\cos\theta_l} d\cos\theta_l} \ . \tag{6}$$

Using the above expressions, we have formulated the following form as:

$$A_{FB}(q^2) = \frac{3I_6^s(q^2)}{\left(3I_1^c(q^2) + 6I_1^s(q^2) - I_2^c(q^2) - 2I_2^s(q^2)\right)} \ . \tag{7}$$

The longitudinal polarization fraction can be extracted as,

$$F_L = \frac{3I_1^c(q^2) - I_2^c(q^2)}{\left(3I_1^c(q^2) + 6I_1^s(q^2) - I_2^c(q^2) - 2I_2^s(q^2)\right)} \ . \tag{8}$$

Here, angular coefficients are expressed in terms of transversity amplitudes $A_\perp^{L,(R)}$, $A_\parallel^{L,(R)}$, $A_0^{L,(R)}$ and $A_t^{L,(R)}$. The form factors are taken from ref. [42] and the detail is discussed in Appendix A.

The expressions of angular coefficients are [37],

$$I_1^c(q^2) = [|A_0^L|^2 + |A_0^R|^2]\left\{(q^4 - (m_1^2 - m_2^2)^2)/q^4\right\} + \frac{(8m_1 m_2)}{q^2}[Re(A_0^L A_0^{R*} - A_t^L A_t^{R*})]$$
$$- 2\frac{(m_1^2 - m_2^2)^2 - q^2(m_1^2 + m_2^2)}{q^4}\{|A_t^L|^2 + |A_t^R|^2\},$$

$$I_1^s(q^2) = \left(|A_\parallel^R|^2 + |A_\perp^R|^2 + \{R \leftrightarrow L\}\right)\frac{\lambda_q + 2[q^4 - (m_1^2 - m_2^2)^2]}{4q^4}$$
$$+ \frac{(4m_1 m_2)}{q^2}\left[Re\left(A_\parallel^L A_\parallel^{R*} + A_\perp^L A_\perp^{R*}\right)\right],$$

$$I_2^c(q^2) = -\frac{\lambda_q}{4q^4}[|A_0^L|^2 + |A_0^R|^2],$$

$$I_2^s(q^2) = \frac{\lambda_q}{q^4}\left(|A_\parallel^R|^2 + |A_\perp^R|^2 + \{R \leftrightarrow L\}\right),$$

$$I_3(q^2) = \frac{\lambda_q}{2q^4}\left(|A_\perp^R|^2 - |A_\parallel^R|^2 + \{R \leftrightarrow L\}\right),$$

$$I_4(q^2) = \frac{\lambda_q}{\sqrt{2}q^4}\left(Re\left(A_\parallel^L A_0^{L*} + \{L \leftrightarrow R\}\right)\right),$$

$$I_5(q^2) = \frac{\sqrt{2\lambda_q}}{q^2}\left[\left(Re(A_0^L A_\perp^{L*} - \{L \leftrightarrow R\})\right) - \frac{(m_1^2 - m_2^2)}{q^2}Re\left(A_t^L A_\parallel^{L*} + \{L \leftrightarrow R\}\right)\right],$$

$$I_6^s(q^2) = -\frac{2\sqrt{\lambda_q}}{q^2}\left(Re\left(A_\parallel^L A_\perp^{L*} - \{L \leftrightarrow R\}\right)\right),$$

$$I_6^c(q^2) = -\frac{4\sqrt{\lambda_q}}{q^2}\frac{(m_1^2 - m_2^2)}{q^2}\left(Re(A_0^L A_t^{L*} + \{L \leftrightarrow R\})\right),$$

$$I_7(q^2) = -\frac{\sqrt{2\lambda_q}}{q^2}\left[\left(Im(A_0^L A_\parallel^{L*} - \{L \leftrightarrow R\})\right) + \frac{(m_1^2 - m_2^2)}{q^2}\left(Im(A_0^L A_t^{L*} + \{L \leftrightarrow R\})\right)\right],$$

$$I_8(q^2) = \frac{\sqrt{\lambda_q}}{\sqrt{2}q^4}\left(Im(A_0^L A_\perp^{L*} + \{L \leftrightarrow R\})\right),$$

$$I_9(q^2) = \frac{\lambda_q}{q^4}\left(Im\left(A_\perp^L A_\parallel^{L*} + \{L \leftrightarrow R\}\right)\right). \tag{9}$$

The expressions of transversity amplitudes are

$$A_\perp^{L,(R)} = \sqrt{2}N_V\sqrt{\lambda_B}\left[(C_9' \mp C_{10}')\frac{V(q^2)}{(m_B + m_V)}\right],$$

$$A_\parallel^{L,(R)} = -\sqrt{2}N_V(m_B^2 - m_V^2)\left[(C_9' \mp C_{10}')\frac{A_1(q^2)}{(m_B - m_V)}\right],$$

$$A_0^{L,(R)} = -\frac{N_V}{2m_V\sqrt{q^2}}\left[(C_9' \mp C_{10}')\left\{(m_B^2 - m_V^2 - q^2)(m_B + m_V)A_1(q^2) - \frac{\lambda_B A_2(q^2)}{m_B + m_V}\right\}\right],$$

$$A_t^{L,(R)} = -\frac{N_V\sqrt{\lambda_B}}{\sqrt{q^2}}\left[(C_9' \mp C_{10}')A_0(q^2)\right]. \tag{10}$$

Here $\lambda$ is the triangular function. $\lambda_B = \lambda(m_B, m_V, \sqrt{q^2})$ and $\lambda_q = \lambda(m_1, m_2, \sqrt{q^2})$ and $N_V = G_F \alpha V_{tb} V_{ts}^* \left(\tau_B \frac{\sqrt{\lambda_B \lambda_q}}{3 \times 2^{10} m_B^3 \pi^5}\right)^{1/2}$.

## IV.   $Z'$ contribution in $B \to K^*(\varphi)l_1^- l_2^+$ decays

In the non-universal $Z'$ model, the newly announced massive gauge boson $Z'$ couples to quarks and the lepton pair. Here, an additional $U(1)'$ gauge group is introduced with the gauge group of the SM [23, 43]. The FCNC transitions are successfully illustrated at tree level by the characteristic of the $Z'$ couplings with fermions. According to refs. [44-46] $Z'$ boson associates with the third generation of quark differently from the other two generations and it is similar for the lepton families also. We have considered different family non-universal $Z'$ couplings for different families in the model. In this work, we have studied with the diagonal and non-universal $Z'$ couplings.

Including $U(1)'$ currents in the BSM physics we can write,

$$J_\mu = \sum_{i,j} \bar{\psi}_j \gamma_\mu \left[\epsilon_{\psi_{L_{ij}}} P_L + \epsilon_{\psi_{R_{ij}}} P_R\right] \psi_i. \tag{11}$$

Here, this sum is done over all fermions and $\psi_{i,j}$ and $\epsilon_{\psi_{R,L_{ij}}}$ are the chiral couplings of the new gauge boson. Actually the FCNCs originate in both left-handed and right-handed sectors at the tree level. Therefore, we can represent as, $B_{ij}^{\psi_L} \equiv \left(V_L^\psi \epsilon_{\psi_L} V_L^{\psi\dagger}\right)_{ij}$ and $B_{ij}^{\psi_R} \equiv \left(V_R^\psi \epsilon_{\psi_R} V_R^{\psi\dagger}\right)_{ij}$. The flavour changing quark coupling $Z'\bar{b}s$ can be generated as

$$\mathcal{L}_{FCNC}^{Z'} = -g'\left(B_{sb}^L \bar{s}_L \gamma_\mu b_L + B_{sb}^R \bar{s}_R \gamma_\mu b_R\right)Z'^\mu + h.c., \tag{12}$$

where $g'$ is the new gauge coupling linked with the extra $U(1)'$ group and the effective Hamiltonian has the form as,

$$H_{eff}^{Z'} = \frac{8G_F}{\sqrt{2}} \left( \rho_{sb}^L \bar{s}_L \gamma_\mu b_L + \rho_{sb}^R \bar{s}_R \gamma_\mu b_R \right) \left( \rho_{l_i l_j}^L \bar{l}_{j_L} \gamma_\mu l_{i_L} + \rho_{l_i l_j}^R \bar{l}_{j_R} \gamma_\mu l_{i_R} \right), \qquad (13)$$

where $\rho_{l_i l_j}^{L,R} \equiv \frac{g' M_Z}{g M_{Z'}} B_{l_i l_j}^{L,R}$, $g$ and $g'$ are the gauge couplings of $Z$ and $Z'$ bosons (here, $g = \frac{e}{\sin\theta_W \cos\theta_W}$) respectively. Here, we have introduced some simplifications which are: (i) we have ignored kinetic mixing, (ii) we have also neglected the mixing of $Z - Z'$ [23, 47-50] (the upper bound of mixing angle is found $10^{-3}$ by Bandyopadhyay et al. [32] and $10^{-4}$ by Bobovnikov et al. [51]) for its very small value, (iii) there are no significant contributions of renormalization group (RG) evolution between $M_W$ and $M_{Z'}$ scales, (iv) we accept the considerable contribution of the flavour-off-diagonal left-handed quark couplings in the flavour changing quark transition [52-56] in our investigation. The detail of this assumption is discussed in Appendix C. The value of $\left|\frac{g'}{g}\right|$ is not fixed yet. But we have considered $\left|\frac{g'}{g}\right| \sim 1$ as both $U(1)$ gauge groups included in our model generate from the same origin of some GUT and $\left|\frac{M_Z}{M_{Z'}}\right| \sim 0.1$ for $Z'$ of TeV-scale . The LEP experiments have also recommended the $Z'$ existence with the identical couplings as the SM $Z$ boson. If $|\rho_{sb}^L| \sim |V_{tb} V_{ts}^*|$, then $B_{sb}^L$ will be $\mathcal{O}(10^{-3})$. With the above summarization the effective Hamiltonian for $b \to s l_1^- l_2^+$ is structured as

$$H_{eff}^{Z'} = -\frac{2G_F}{\sqrt{2}\pi} V_{tb} V_{ts}^* \left[ \frac{B_{sb}^L B_{l_1 l_2}^L}{V_{tb} V_{ts}^*} (\bar{s}b)_{V-A} (\bar{l}_1 l_2)_{V-A} - \frac{B_{sb}^L B_{l_1 l_2}^R}{V_{tb} V_{ts}^*} (\bar{s}b)_{V-A} (\bar{l}_1 l_2)_{V+A} \right] + h.c, (14)$$

where $B_{sb}^L$ is the left-handed coupling of $Z'$ boson with quarks, $B_{l_1 l_2}^L$ and $B_{l_1 l_2}^R$ are the left-handed and right-handed couplings with the leptons respectively. The coupling parameter consists of a NP weak phase term, which is related as $B_{sb}^L = |B_{sb}^L| e^{-i\varphi_s^l}$.

Considering eq. (1) and eq. (2), we have incorporated the NP terms as follows:

$$C_9^{NP} = \frac{4\pi B_{sb}^L}{\alpha V_{tb} V_{ts}^*} \left( B_{l_1 l_2}^L + B_{l_1 l_2}^R \right),$$

$$C_{10}^{NP} = \frac{4\pi B_{sb}^L}{\alpha V_{tb} V_{ts}^*} \left( B_{l_1 l_2}^L - B_{l_1 l_2}^R \right). \qquad (15)$$

Including the NP parts through the modified Wilson coefficients, we have studied the observables defined in eqs. (5), (7) and (8) in the BSM physics.

### V.   Numerical Analysis

In this work, we study the differential branching ratio, forward backward asymmetry and longitudinal polarization for $b \to s$ transition. In our previous work [57], we have studied lepton flavour violating $\Lambda_b$ decays and constrained the LFV leptonic couplings from the recent bound of various LFV decays of same quark transition. The LFV quark couplings are taken from refs [39, 58] and are recorded in Table-1. The form factor values are taken from Appendix B [42]. The values of other input parameters are tabulated in the Appendix D.

**Table-1: Numerical values of coupling parameters**

| Scenarios | $|B_{sb}| \times 10^{-3}$ | $\varphi_s^l$ (in degree) |
|---|---|---|
| $S_1$ | $(1.09 \pm 0.22)$ | $(-72 \pm 7)°$ |
| $S_2$ | $(2.20 \pm 0.15)$ | $(-82 \pm 4)°$ |

Here, we have considered the maximised values of the coupling parameters to intensify the NP in these decays. The leptonic couplings are composed in the following way:

$$S_{l_1 l_2} = \left(B_{l_1 l_2}^L + B_{l_1 l_2}^R\right), \text{ and } D_{l_1 l_2} = \left(B_{l_1 l_2}^L - B_{l_1 l_2}^R\right). \tag{16}$$

The scenario $C_9' = C_{10}'$ is strongly abandoned because it produces the result $R_K \gtrsim 1$ [62] which is not in agreement with the experiments. In ref. [63] also this consideration $C_9' = C_{10}'$ does not match the experimental values of branching ratios of $B_s \to \mu^+\mu^-$ and $B \to K\mu^+\mu^-$ at high $q^2$ region; whereas there are plethora of best results of the anomalies with the scenario $C_9' = -C_{10}'$ [40, 59-64]. Not only the model dependent analysis but also the model independent analysis support the fruitfulness of the well-considered scenario. Therefore, we have adapted the scenario $C_9' = -C_{10}'$ in our investigation. Comprising all the above discussions, we have set two scenarios with maximum values of the NP couplings as Table-2 below.

**Table-2: Maximum values of NP couplings**

| Quark couplings | | | Leptonic couplings | | | |
|---|---|---|---|---|---|---|
| Scenarios | $|B_{sb}|$ | $\varphi_s^l$ | | | | |
| $S_1$ | $(1.31 \times 10^{-3})$ | $(-65)°$ | $S_{\mu e}$ | $0.0079$ | $S_{\tau \mu}$ | $0.11$ |
| $S_2$ | $(2.35 \times 10^{-3})$ | $(-78)°$ | $D_{\mu e}$ | $-0.0079$ | $D_{\tau \mu}$ | $-0.11$ |

With all these numerical data we have varied the differential branching ratio within allowed kinematic region of $q^2$ in Fig. 1a and Fig. 1b for $B \to K^* \mu\tau$ and $B \to K^* e\mu$ channels respectively and Fig. 2a and Fig. 2b for $B_s \to \varphi\mu\tau$ and $B_s \to \varphi e\mu$ channels respectively. Here, all the plots are drawn for the central values of the input parameters and form factors. The blue line represents the variation for scenario 2 and orange line represents the scenario 1. We observe that the values of differential branching ratios are greater at low kinematic region. The maximum value at scenario 2 infers the appreciable contribution of NP on the decays. Integrating over the whole $q^2$ phase space we have estimated the average values of branching ratios for these decays which are shown in Table-3. We assume 10% uncertainties on the form factors [42] (as it is one of the main sources of vagueness in the evaluation of observable) and the error on the input values (which are indicated in Appendix D) in our calculation.

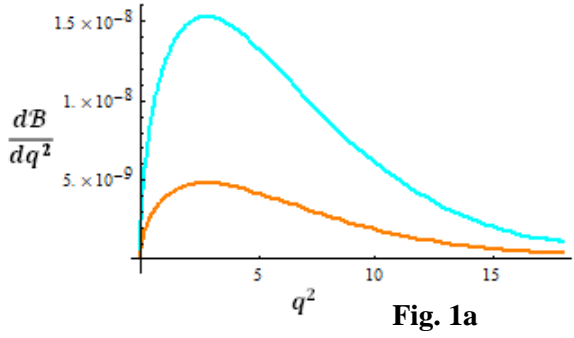
Fig. 1a

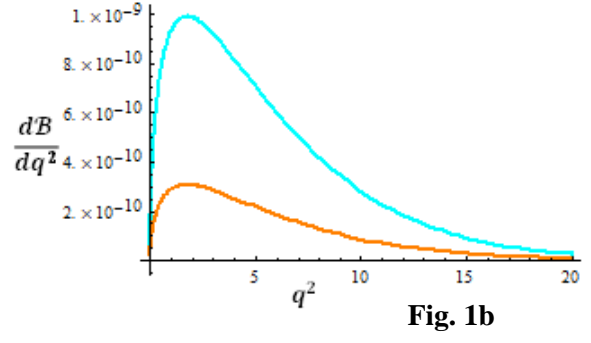
Fig. 1b

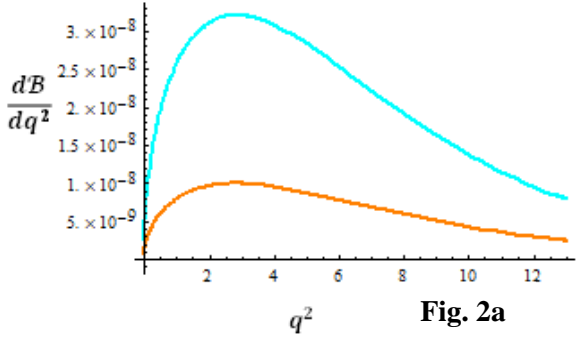
Fig. 2a

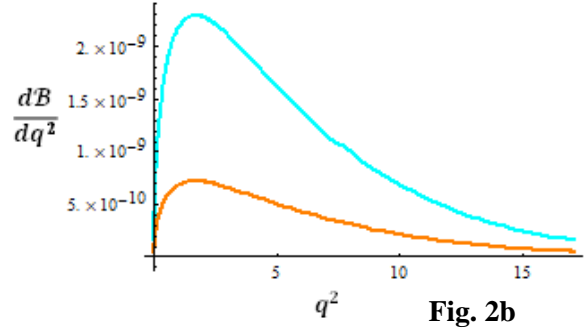
Fig. 2b

**Fig: Variation of differential branching ratio for (1a) $B \to K^* \mu^- \tau^+$ (1b) $B \to K^* e^- \mu^+$ (2a) $B_s \to \varphi \mu^- \tau^+$ (2b) $B_s \to \varphi e^- \mu^+$ within allowed kinematic region of $q^2$ using the bound of NP couplings**

**Table-3: Predicted values of branching ratios for LFV $B \to K^*$ and $B_s \to \varphi$ decays in 1st and 2nd scenarios**

| Decay mode | Branching ratio value | |
|---|---|---|
| | Scenario 1 ($S_1$) | Scenario 2 ($S_2$) |
| $B \to K^* \mu^- \tau^+$ | $(2.11 \pm 1.02) \times 10^{-9}$ | $(9.72 \pm 2.56) \times 10^{-9}$ |
| $B \to K^* e^- \mu^+$ | $(1.16 \pm 0.65) \times 10^{-10}$ | $(5.42 \pm 1.22) \times 10^{-10}$ |
| $B_s \to \varphi \mu^- \tau^+$ | $(6.51 \pm 3.02) \times 10^{-9}$ | $(2.09 \pm 0.82) \times 10^{-8}$ |
| $B_s \to \varphi e^- \mu^+$ | $(3.40 \pm 1.84) \times 10^{-10}$ | $(1.09 \pm 0.32) \times 10^{-9}$ |

Another promising tool to explore NP in these decays is the lepton side forward-backward asymmetries. The sensitiveness of the observable on the LFV $\Lambda_b$ baryon decays has been studied in the references [40, 57]. We have studied the variation of forward-backward asymmetries in the $Z'$ model in the allowed $q^2$ region in Fig. 3a and Fig. 3b for $B \to K^* \tau \mu$ and $B \to K^* \mu e$ channels respectively, whereas Fig. 4a and Fig. 4b represent the variance for $B_s \to \varphi \tau \mu$ and $B_s \to \varphi \mu e$ channels respectively. The colour description of the figures is similar as previous. The plots interpret that the zero crossing positions are very responsive to NP. Here, Fig. 3a shows the position of zero crossing of $B \to K^* \tau \mu$ for scenario 2 at $q^2 = 10.58$ GeV$^2$ and for scenario 1 at $q^2 = 10.08$ GeV$^2$ whereas in Fig. 3b the scenario 2 shows the zero crossing at $q^2 = 6.245$ GeV$^2$ for $B \to K^* \mu e$ but it is absent for scenario 1. The zero crossing is also absent for $B_s \to \varphi \tau \mu$ in Fig. 4a and the observable decreases gradually, reaches at minimum value and then again increases. At low $q^2$ region the variation of both the scenarios are not considerable but at high $q^2$ regime the 2$^{nd}$ scenario attains the minimum value. And Fig. 4b shows the variation of the observable for $B_s \to \varphi\mu e$ channel. Here, we observe the zero crossing for the 2$^{nd}$ scenario only at $q^2 = 1.495$ GeV$^2$ and the 1$^{st}$ scenario has increasing nature throughout the allowed $q^2$ region. The values of the forward-backward asymmetries are calculated and represented in Table-4.

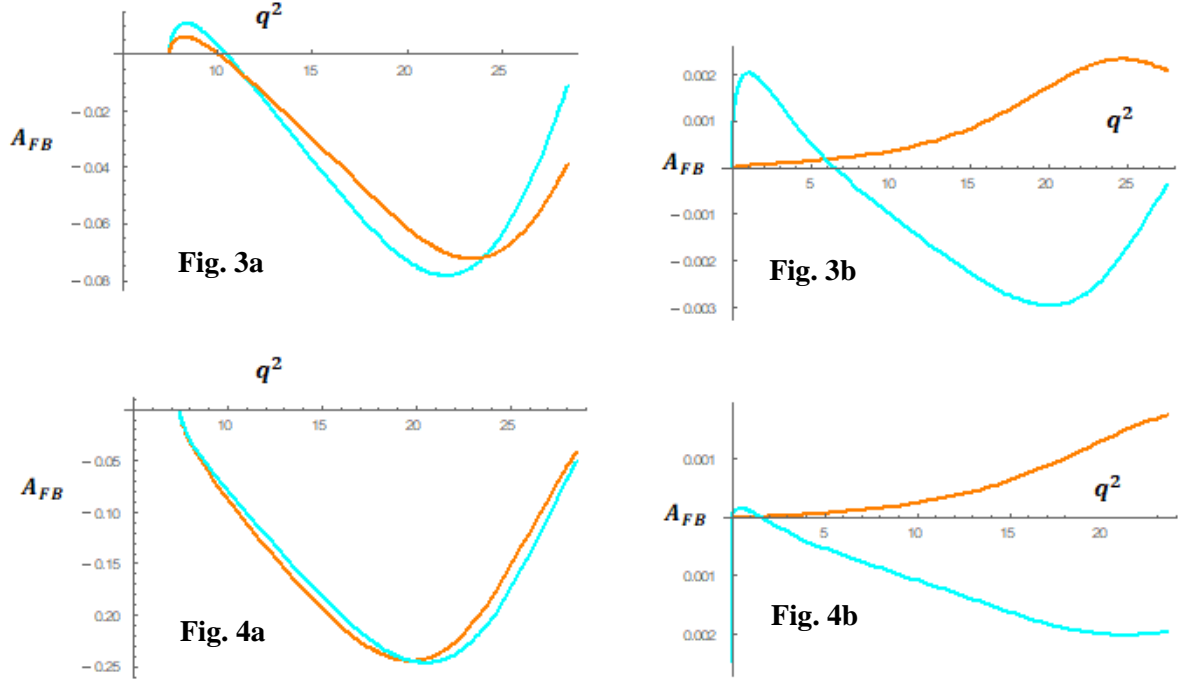

**Fig: Variation of forward-backward asymmetry for (3a) $B \to K^*\mu^-\tau^+$ (3b) $B \to K^*e^-\mu^+$ (4a) $B_s \to \varphi\mu^-\tau^+$ (4b) $B_s \to \varphi e^-\mu^+$ within allowed kinematic region of $q^2$ using the bound of NP couplings**

**Table-4: Predicted values of forward-backward asymmetries for LFV $B \to K^*$ and $B_s \to \varphi$ decays in 1$^{st}$ and 2$^{nd}$ scenarios**

| Decay mode | Forward-backward asymmetry ($A_{FB}$) | |
|---|---|---|
| | Scenario 1 ($S_1$) | Scenario 2 ($S_2$) |
| $B \to K^*\mu^-\tau^+$ | $-0.0302 \pm 0.50$ | $-0.0343 \pm 0.47$ |
| $B \to K^*e^-\mu^+$ | $0.00005 \pm 0.00002$ | $-0.0012 \pm 0.0008$ |
| $B \to \varphi\mu^-\tau^+$ | $-0.1501 \pm 0.0810$ | $-0.1522 \pm 0.0608$ |
| $B \to \varphi e^-\mu^+$ | $0.00035 \pm 0.00018$ | $-0.0009 \pm 0.0005$ |

Longitudinal polarization asymmetry, one of the most efficient tools to test NP, for $B \to K^*(\to K^-\pi^+)l_1 l_2$ and $B_s \to \varphi(\to K\bar{K})l_1 l_2$ decays are studied in non-universal $Z'$ model for the first time in this work. The variation of the observable is presented in the allowed kinematic regime for $B \to K^*\tau\mu$ and $B \to K^*\mu e$ decays in the Fig. 5a and Fig. 5b respectively. Similar variation is shown for $B_s \to \varphi\tau\mu$ and $B_s \to \varphi\mu e$ channels in Fig. 6a and Fig. 6b respectively. The nature of the asymmetries conveys similar manner for all the channels shown above. The values of the observables do not drop at the zero and for $B \to K^*\tau\mu$ decay the 2$^{nd}$ scenario maintains almost constant value throughout the whole $q^2$ region. But it drops comparatively in sharp manner at high $q^2$ region for the other decay modes. The values of longitudinal polarization are tabulated in Table-5.

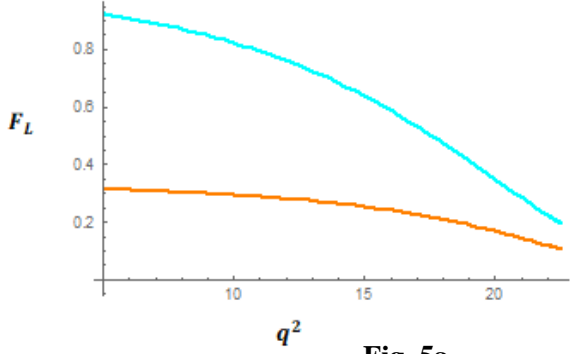
**Fig. 5a**

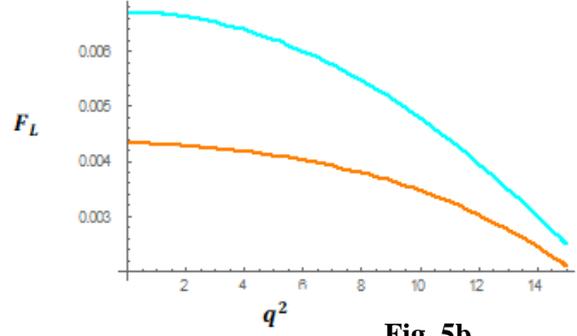
**Fig. 5b**

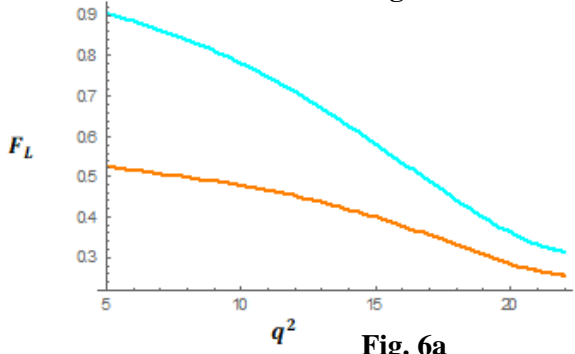
**Fig. 6a**

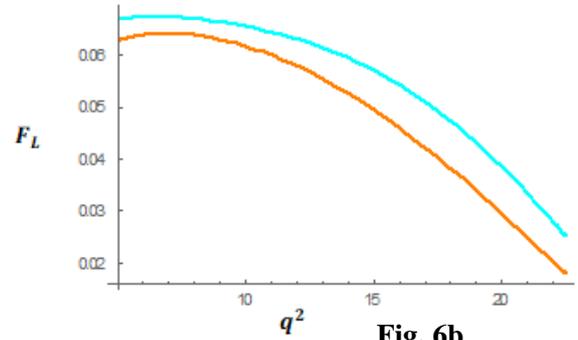
**Fig. 6b**

**Fig: Variation of longitudinal polarization fraction for (5a) $B \to K^*\mu^-\tau^+$ (5b) $B \to K^*e^-\mu^+$ (6a) $B_s \to \varphi\mu^-\tau^+$ (6b) $B_s \to \varphi e^-\mu^+$ within allowed kinematic region of $q^2$ using the bound of NP couplings**

**Table-5: Predicted values of longitudinal polarization fractions for LFV $B \to K^*$ and $B_s \to \varphi$ decays in 1$^{st}$ and 2$^{nd}$ scenarios**

| Decay mode | Longitudinal Polarization ($F_L$) | |
|---|---|---|
| | Scenario 1 ($S_1$) | Scenario 2 ($S_2$) |
| $B \to K^*\mu^-\tau^+$ | $0.2787 \pm 0.1910$ | $0.6208 \pm 0.1510$ |
| $B \to K^*e^-\mu^+$ | $0.0035 \pm 0.0018$ | $0.0052 \pm 0.0022$ |
| $B_s \to \varphi\mu^-\tau^+$ | $0.3907 \pm 0.2420$ | $0.7113 \pm 0.2050$ |
| $B_s \to \varphi e^-\mu^+$ | $0.0500 \pm 0.0263$ | $0.0561 \pm 0.0102$ |

## VI. Conclusion

Undoubtedly the SM achieved an enormous success in the theoretical community of high energy physics. But lepton flavour violation is still suppressed in the SM which infers its imperfection. So it can be said that there must be some physics beyond the SM. Recently the LHCb has set a few bounds on LFV $b \to s l_1^- l_2^+$ decays [65]. These results of LHCb accelerate the BSM physics towards a new horizon incorporating several advanced and remarkable analysis. Previously the LFV decays are studied in different NP models. In this paper, we have studied $B \to K^* l_1^- l_2^+$ and $B_s \to \varphi l_1^- l_2^+$ decays in non-universal $Z'$ model. We have calculated the branching ratio values, forward backward asymmetries and longitudinal polarization asymmetries for these decays. In our last work, we have constrained the LFV leptonic couplings [57] which are applied here. Our plots show that the 2$^\text{nd}$ scenario provides higher values of the observables which offer a clear cut signature of NP in these LFV decays. We notice the existence of zero crossings in the plots of $A_{FB}$ but $B \to K^* e\mu$ and $B_s \to \varphi \mu\tau$ decays do not show the zero crossing positions. Actually zero crossing position is the point where the uncertainties due to form factor terms are cancelled out and the sensitivity due to NP can be identified. The zero position changes for two different scenarios of NP in Fig. 3a, Fig. 3b and Fig. 4a and this responsiveness proves the effect of NP. The $A_{FB}$ value attains greater negative value for $B_s \to \varphi \mu\tau$ decay whereas $B(B_s) \to K^*(\varphi) e\mu$ have positive value throughout the whole kinematic region. These changes in nature of the observable are due to leptonic masses. Another interesting observable we have included in this work is longitudinal polarization fraction. In the advance 2000's, this observable has gained attention for the first time through the well-known Polarization Puzzle. We observe that the observable serves almost similar variation for all the decays and also the higher value of NP couplings provide higher value of the observable. This character evolves the probing nature of the observable towards NP.

Previously the lepton violating B decays are studied with the effect of scalar leptoquarks [66], in SUSY [67], with the contribution of pseudo-scalar co-efficients (coupling to Higgs) [37], in generic $Z'$ models [38]. These previous studies are mainly based upon the branching ratio except the reference [37]. In scalar leptoquark model and SUSY the upper limit of the branching ratio for $B \to K^* e\mu$ was predicted as $(3.4 \times 10^{-7})$ whereas we have calculated as $(1.16 \pm 0.65) \times 10^{-10}$ for scenario 1, $(5.42 \pm 1.22) \times 10^{-10}$ for scenario 2 and experimentally it is set at $(1.2 \times 10^{-7})$ with 90% C. L. On the other hand, the branching ratio for $B \to K^* \mu\tau$ decay was predicted as $(8.21 \times 10^{-5})$ in leptoquark model, $(11.3 \times 10^{-5})$ in SUSY, $(1.6 \times 10^{-8})$ in generic $Z'$ models, $(9.3 \times 10^{-8})$ with the contribution of pseudo-scalar co-efficients and we have estimated as $(2.11 \pm 1.02) \times 10^{-9}$ for scenario 1, $(9.72 \pm 2.56) \times 10^{-9}$ for scenario 2. The $B_s \to \varphi l_1^- l_2^+$ decays are not studied experimentally in B factories till now. The asymmetries of the relevant B decays studied in this paper are also beyond the colliders. Here these parameters are studied in the non-universal $Z'$ model for the first time.

With the quick expansion of LHCb energy and upcoming Belle II experiments the LFV decays can be investigated with high precision and accuracy. The study of these decays are actually smoking signal of NP which will be proved in near future with the developments of colliders.

The results obtained in Table-3, 4 and 5 will be very useful to the experimentalists as well as the theorists.

**Acknowledgement**

We thank the reviewer for suggesting valuable improvements of our manuscript. S. Biswas and S. Mahata thank National Institute of Technology Durgapur for providing fellowship during their research. A. Biswas and S. Sahoo acknowledge the SERB, DST, Govt. of India for financial assistance through project (TAR/2021/000036).

**Appendix A**

The hadronic matrix elements can be parametrized in terms of various form factors as defined below

$$\langle V(k)|\bar{s}\gamma^{\mu}(1-\gamma_5)b|\bar{B}(p)\rangle$$
$$= \varepsilon_{\mu\nu\rho\sigma}\varepsilon^{*\nu}p^{\rho}k^{\sigma}\frac{2V(q^2)}{m_B+m_V} - i\varepsilon_{\mu}^{*}(m_B+m_V)A_1(q^2)$$
$$+ i(p+k)_{\mu}(\varepsilon^{*}.q)\frac{A_2(q^2)}{m_B+m_V}$$
$$+ iq_{\mu}(\varepsilon^{*}.q)\frac{2m_V}{q^2}[A_3(q^2) - A_0(q^2)], \quad (A1)$$

$$\langle V(k)|\bar{s}\sigma_{\mu\nu}q^{\nu}(1-\gamma_5)b|\bar{B}(p)\rangle$$
$$= 2i\varepsilon_{\mu\nu\rho\sigma}\varepsilon^{*\nu}p^{\rho}k^{\sigma}T_1(q^2) + [\varepsilon_{\mu}^{*}(m_B^2-m_V^2) - (2p-q)_{\mu}(\varepsilon^{*}.q)]T_2(q^2)$$
$$+ (\varepsilon^{*}.q)\left[q_{\mu} - \frac{q^2}{m_B^2-m_V^2}(p+k)_{\mu}\right]T_3(q^2), \quad (A2)$$

where $\varepsilon_{\mu}$ is the polarization vector of the vector meson $K^*(\varphi)$. Among the form factors, $A_3(q^2)$ is a dependent quantity as, $2m_V A_3(q^2) = (m_B+m_V)A_1(q^2) - (m_B-m_V)A_2(q^2)$.

**Appendix B**

According to reference [42], we have structured the form factors as below,

$$F(q^2) = \frac{r_1}{1-\frac{q^2}{m_{fit}^2}} + \frac{r_2}{\left(1-\frac{q^2}{m_{fit}^2}\right)^2}, \quad (B1)$$

The central values of form factors for $B \to K^*(\to K^-\pi^+)l_1l_2$ and $B_s \to \varphi(\to K\bar{K})l_1l_2$ decays are recorded in Table-6 and Table-7.

**Table-6: Form factors for $B \to K^* l_1 l_2$**

| Form Factors | $r_1$ | $r_2$ | $m_{fit}^2$ |
|---|---|---|---|
| V | 0.923 | −0.511 | 49.40 |
| $A_0$ | 1.364 | −0.990 | 36.78 |
| $A_1$ | — | 0.290 | 40.38 |
| $A_2$ | −0.084 | 0.342 | 52.00 |
| $T_1$ | 0.823 | −0.491 | 46.31 |
| $T_2$ | — | 0.333 | 41.41 |
| $T_3$ | −0.036 | 0.368 | 48.10 |

**Table-7: Form factors for $B \to \varphi l_1 l_2$**

| Form Factors | $r_1$ | $r_2$ | $m_{fit}^2$ |
|---|---|---|---|
| V | 1.484 | −1.049 | 39.52 |
| $A_0$ | 3.310 | −2.835 | 31.57 |
| $A_1$ | — | 0.308 | 36.54 |
| $A_2$ | −0.054 | 0.288 | 48.94 |
| $T_1$ | 1.303 | −0.491 | 46.31 |
| $T_2$ | — | 0.333 | 41.41 |
| $T_3$ | 0.027 | 0.321 | 45.56 |

**Appendix C**

The new $Z'$ couplings help to explain the FCNC transitions at tree level. The FCNCs depend on the individual unitary transformations for the left and right chiral quarks (u and d), leptons and neutrinos diagonalizing their particular mass matrices [23]. However, only the combination of left chiral quarks (u and d) that are occurred in the CKM matrix is known experimentally (with weak constraints on the leptonic analog from neutrino oscillations).

Mathematically, we have the current correspond to the additional $U(1)$ gauge group in eq. (11) where $\epsilon_{\psi_{R,L_{ij}}}$ are the right handed and left handed chiral couplings. These chirals for up quark are flavour diagonal and family universal as [54-56]: $\epsilon_{L,R}^u = Q_{L,R}^u I$, $\epsilon_{L,R}^e = Q_{L,R}^e I$ and $\epsilon_L^\nu = Q_L^\nu I$ where $I$ represents the $3 \times 3$ identity matrix and $Q_{L,R}^u$, $Q_{L,R}^e$ and $Q_L^\nu$ represent the chiral charges. The chirals for down quark are represented as:

$$\epsilon_L^d = Q_L^d \begin{pmatrix} 1 & 0 & 0 \\ 0 & 1 & 0 \\ 0 & 0 & x \end{pmatrix} \text{ and } \epsilon_R^d = Q_R^d \begin{pmatrix} 1 & 0 & 0 \\ 0 & 1 & 0 \\ 0 & 0 & 1 \end{pmatrix}.$$ Here, the family non-universality can be signalized by the term $x$. Diagonalization of Yukawa matrices indicates the right-handed

couplings as flavour-diagonal whereas the non-diagonal left handed part unveil the non-universal nature of the flavour couplings. So in our work, we have considered the left-handed $b - s - Z'$ couplings and neglected the couplings of the $Z'$ with right-handed quarks.

**Appendix D:**

**Table-8: Values of other input parameters [64]**

| Parameter | Values |
|---|---|
| $m_\mu$ | $(105.66 \pm 0.0000024)$ MeV |
| $m_e$ | $(0.51 \pm 0.0000000031)$ MeV |
| $m_\tau$ | $(1776.86 \pm 0.12)$ MeV |
| $m_{K^*}$ | $(891.66 \pm 0.26)$ MeV |
| $m_\varphi$ | $(1019.461 \pm 0.016)$ MeV |
| $m_B$ | $(5279.55 \pm 0.26)$ MeV |
| $m_{B_s}$ | $(5366.84 \pm 0.15)$ MeV |
| $m_\pi$ | $(139.57039 \pm 0.00017)$ MeV |
| $m_K$ | $(497.611 \pm 0.013)$ MeV |
| $G_F$ | $(1.166 \pm 0.0000006) \times 10^{-5}$ GeV$^{-2}$ |
| $|V_{tb}|$ | $(1.019 \pm 0.025)$ |
| $|V_{ts}|$ | $(39.4 \pm 2.3) \times 10^{-3}$ |